\documentclass[twocolumn,showpacs,preprintnumbers,amsmath,amssymb,prl,footinbib,bibnotes,superscriptaddress]{revtex4}
\usepackage{times}
\usepackage{graphicx}
\usepackage{cases}
\newcommand{\abs}[1]{\lvert#1\rvert}
\begin{document}
\bibliographystyle{apsrev}

\title{Existence and Vanishing of the Breathing Mode in Strongly Correlated Finite Systems}
\author{C.~Henning}
\affiliation{Institut f\"ur Theoretische Physik und Astrophysik, Christian-Albrechts-Universit\"{a}t zu Kiel, D-24098 Kiel, Germany}
\author{K. Fujioka}
\affiliation{The City College of New York, New York City, NY 10031, USA}
\author{P.~Ludwig}
\affiliation{Institut f\"ur Theoretische Physik und Astrophysik, Christian-Albrechts-Universit\"{a}t zu Kiel, D-24098 Kiel, Germany}
\author{A.~Piel}
\affiliation{Institut f\"ur Experimentelle und Angewandte Physik, Christian-Albrechts-Universit\"{a}t zu Kiel, D-24098 Kiel, Germany}
\author{A.~Melzer}
\affiliation{Institut f\"ur Physik, Universit\"{a}t Greifswald, D-17489 Greifswald, Germany}
\author{M.~Bonitz}
\email{bonitz@physik.uni-kiel.de}
\affiliation{Institut f\"ur Theoretische Physik und Astrophysik, Christian-Albrechts-Universit\"{a}t zu Kiel, D-24098 Kiel, Germany}

\date{\today}

\begin{abstract}
One of the fundamental eigenmodes of finite interacting systems is the mode of {\em uniform radial expansion and contraction} -- the ``breathing'' mode (BM). Here we show in a general way that this mode exists only under special conditions: i) for harmonically trapped systems with interaction potentials of the form $1/r^\gamma$ $(\gamma\in\mathbb{R}_{\neq0})$ or $\log(r)$, or ii) for some systems with special symmetry such as single shell systems forming platonic bodies. Deviations from the BM are demonstrated for two examples: clusters interacting with a Lennard-Jones potential and parabolically trapped systems with Yukawa repulsion. We also show that vanishing of the BM leads to the occurence of multiple monopole oscillations which is of importance for experiments.
\end{abstract}
\maketitle

Strongly correlated finite systems are of high current interest in many fields. Examples are atomic clusters, e.g., \cite{baletto05,proykova06}, and externally confined systems interacting by repulsive potentials, such as atomic Bose condensates or strongly interacting Fermi gases \cite{moritz03,kinast_pra04}, ion crystals in traps \cite{itano}, dusty plasma crystals, e.g., \cite{arp04,bonitz-etal.prl06}, or electrons in quantum dots \cite{afilinov-etal.prl01,reimann02}.
The response of these systems to small external pertubations is fully determined by its collective excitations. Among those, the radial expansion and contraction - the monopole oscillation (MO) - is particularly important since it can be easily excited selectively by variation of the confinement \cite{melzer.prl01} or by applying external fields \cite{dykeman08}. The corresponding frequency $\omega_\textrm{MO}$ can often be precisely measured and may serve as a sensitive indicator of intrinsic system properties including the form of the pair interaction, the trap geometry \cite{kinast_pra04} or the screening length and particle charge in complex plasmas \cite{melzer.prl01}.

On the theory side, the collective modes are successfully analyzed within continuum models, e.g., \cite{dubin_schiffer.pre96,sheridan.physplasmas06}. These models are applicable to the gas or fluid phases of classical or quantum systems where correlations are weak or moderate. In the strongly correlated crystalline state where the individual particle positions ${\bf r}_i$ become separated, such models are questionable. Nevertheless, frequently a MO is associated with the oscillation of the mean square radius, ${\bf R}^2(t)=N^{-1}\sum_i {\bf r}_i^2(t)$, \cite{schweigert_etal.prb95,partoens_etal.jphy97,melzer.prl01}. In three-dimensional (3D) systems with Coulomb interaction and harmonic confinement with frequency $\omega_\textrm{0}$ this MO has the frequency $\omega_\textrm{MO}=\sqrt{3}\,\omega_\textrm{0}$, independent of $N$, e.g., \cite{schweigert_etal.prb95,dubin_schiffer.pre96}. In harmonically confined 2D systems, also a universal ($N$-independent) MO was observed if the interaction is a repulsive power law,  $\sim1/r^n\,(n=1,2,\ldots)$, or logarithmic \cite{partoens_etal.jphy97}. For other interactions, $\omega_\textrm{MO}$ is $N$-dependent \cite{calvo.epjd07}.

On the other hand, the response of such a strongly correlated finite system to small pertubations is fully determined by its normal modes. Among those, the mode of {\em uniform radial expansion and contraction}, the breathing mode (BM), is most similar to the MO and both are often used synonymously.

This common identification of the breathing mode with the monopole oscillation assumes that {\em a BM exists always}, independently of the confinement and for any pair interaction. In this Letter we show, by a general direct analytical investigation, that this assumption does not hold for strongly correlated finite classical systems.
We prove that in fact a {\em BM exists only} in two classes of systems: i) in parabolically trapped systems of any dimension with power law or logarithmic pair interaction, and ii) in some highly symmetric systems, for arbitrary pair interaction. In all other cases, {\em any normal mode deviates from the BM}. As a consequence, then there exist several monopole oscillations with different frequencies.
%

{\bf General existence conditions of the BM.}
We consider a $d$-dimensional, classical system of $N$ identical particles with arbitrary pair interaction $v(r)$ described by the hamiltonian
\begin{equation}\label{hamiltonian}
	H=\sum_{i=1}^N\frac{m}{2}\dot{\bf r}^2_i +\underbrace{\sum_{i=1}^N\phi(\abs{{\bf r}_i})
	+\frac{1}{2}\sum_{\substack{i,j=1\\j\neq i}}^N v(\abs{{\bf r}_{ij}})}_{U({\bf r})\textrm{ with }{\bf r}\,\in\mathbb{R}^{dN}}.
\end{equation}
The included confinement potential $\phi$ is isotropic but completely general otherwise and may also be equal to zero. We assume the existence of a stable configuration (ground state or metastable state) ${\bf r}^*=\bigl({\bf r}^*_1,{\bf r}^*_2\ldots{\bf r}^*_N\bigr)\,\in\mathbb{R}^{dN}$ defined by ($\sum'$ indicates $l\neq i$)
\begin{equation}\label{equilibrium_positions}
	0=\nabla_i U({\bf r})\rvert_{{\bf r}={\bf r}^*}=\frac{\phi'(\abs{{\bf r}_i^*})}{\abs{{\bf r}_i^*}}{\bf r}_i^*+\sideset{}{'}\sum_{\substack{l=1}}^{N}\frac{v'(\abs{{\bf r}_{il}^*})}{\abs{{\bf r}_{il}^*}}{\bf r}_{il}^*,
\end{equation}
from where the normal modes are excited. For a general normal mode analysis we use the eigenvalue equation
\begin{equation}\label{EV_matrixform}
	\lambda m \Hat{\bf r}=\mathcal{H}\Hat{\bf r},
\end{equation}
which contains the positive semidefinite Hessian matrix $\mathcal{H}=\left(\nabla_i \nabla_j U({\bf r})\rvert_{{\bf r}={\bf r}^*}\right)_{i,j=1\ldots N}$. Accordingly, $\lambda=\omega^2\geq0$ is the eigenvalue connected with the mode frequency $\omega$, and $\Hat{\bf r}$ is the eigenvector containing the displacement vectors of all particles within the mode.

To obtain the existence conditions of the BM, we first rearrange Eq.~\eqref{EV_matrixform} by evaluating the Hessian matrix $\mathcal{H}$, using the isotropy of $\phi$ and the distance dependence of $v$, which yields, for each component 
$i \in 1\ldots N$
\begin{equation}
\begin{split}
	&\lambda m \Hat{\bf r}_i=\sum_{j=1}^N\mathcal{H}_{ij}\Hat{\bf r}_j\\
	&\quad=\frac{\phi'(\abs{{\bf r}_i^*})}{\abs{{\bf r}_i^*}}\Hat{\bf r}_i
		+\frac{({\bf r}_i^*\cdot\Hat{\bf r}_i){\bf r}_i^*}{\abs{{\bf r}_i^*}^3}
	\Bigl(\abs{{\bf r}_i^*}\phi''(\abs{{\bf r}_i^*})-\phi'(\abs{{\bf r}_i^*})\Bigr)\\
	&\qquad+\sideset{}{'}\sum_{\substack{l=1}}^N
	\biggl[\frac{v'(\abs{{\bf r}_{il}^*})}{\abs{{\bf r}_{il}^*}}\Hat{\bf r}_{il}\\
	&\qquad\qquad\quad+\frac{({\bf r}_{il}^*\cdot\Hat{\bf r}_{il}){\bf r}_{il}^*}{\abs{{\bf r}_{il}^*}^3}
	\Bigl(\abs{{\bf r}_{il}^*}v''(\abs{{\bf r}_{il}^*})-v'(\abs{{\bf r}_{il}^*})\Bigr)\biggr].\raisetag{55pt}
\end{split}
\end{equation}
Now we insert an eigenvector of the BM given by $\Hat{\bf r} \propto {\bf r}^*$ (the proportionality constant cancels) and obtain
\begin{equation}\label{existing_condition}
	\lambda m {\bf r}_i^*=\phi''(\abs{{\bf r}_i^*}){\bf r}_i^*+\sideset{}{'}\sum_{\substack{l=1}}^N v''(\abs{{\bf r}_{il}^*}){\bf r}_{il}^*.
\end{equation}
These are the $N$ existence equations of the BM, which just have the meaning that, the linearized force on each particle from the confinement and the pair interactions has to be purely {\em radial and uniform} (proportional to ${\bf r}_i^*$). To find their solutions we separate them into the two conditions. First, radiality is equivalent to
\begin{equation}\label{existing_condition_normal_component}
	0={\bf r}_i^*\times\sideset{}{'}\sum_{\substack{l=1}}^N v''(\abs{{\bf r}_{il}^*}){\bf r}_{il}^*,
\end{equation}
which requires $\sum_{l}' v''(\abs{{\bf r}_{il}^*}){\bf r}_{il}^*=s_i{\bf r}_i^*$ with arbitrary coefficients $s_i$. These radial equations have two classes of solutions: 
R1) the particle  configuration is of special symmetry with coefficients $s_i$ determined by the configuration, e.g., rotational symmetry with respect to all ${\bf r}_{i}^*$. 
R2) the pair interaction $v(r)$ is of specific form. This form can be obtained by adding a multiple of the tangential components of Eqs.~\eqref{equilibrium_positions} to Eqs.~\eqref{existing_condition_normal_component}. The result is an equation for $v$: $v''(r)=-(\gamma+1)v'(r)/r$, where $\gamma$ is an arbitrary real number. Its solutions are $v(r)\sim 1/r^\gamma$ [$\gamma\neq0$] and  $v(r)\sim \log(r)$ [corresponding to $\gamma=0$], and these solutions fulfill Eqs.~\eqref{existing_condition_normal_component} independently of the particle configuration with coefficients given by $s_i=(\gamma+1)\phi'(\abs{{\bf r}_i^*})/\abs{{\bf r}_i^*}$.

So far we used only the requirement of radiality. To fulfill the existence equations \eqref{existing_condition} the radial solutions R1) and R2) also have to fulfill the conditions of uniformity which follow by multiplying \eqref{existing_condition} with ${\bf r}_i^*$
\begin{equation}\label{existing_condition_radial_component}
	\lambda m=\phi''(\abs{{\bf r}_i^*})+s_i.
\end{equation}
These conditions also have two kinds of solutions: 
U1) the configuration has another special symmetry, e.g., single shell configuration with $s_i$ independent of $i$. 
U2) in case of the radial solution R2) a configuration-independent equation for the confinement potential $\phi(r)$ is obtained by using the corresponding coefficients $s_i$: $\lambda m=\phi''(r)+(\gamma+1)\phi'(r)/r$. Its solutions are
\begin{subequations}
\begin{numcases}{\phi(r)=}
	\frac{\lambda m}{2(2+\gamma)}r^2+c\,v(r)\label{a}& $\gamma\neq-2$ \\[2ex]
	\frac{\lambda m}{2}r^2\log(r)+c\,r^2\label{b}& $\gamma=-2$,
\end{numcases}
\end{subequations}
where $\gamma$ is the power of the pair interaction and $c$ is an arbitrary constant. Consequently, there are two basically different possibilities for the existence of a BM:

{\bf Universal breathing mode (UBM).} 
The previous analysis shows that a configuration- and $N$-independent {\em universal} BM exists in the case of harmonically confined systems, Eq.~\eqref{a}, $\phi(r)=m\omega_\textrm{0}^2r^2/2$, with particles interacting via potentials $v(r)$ proportional to $1/r^\gamma$ or to $\log(r)$. For these cases the breathing frequency is independent of $N$ and given by $\omega_\textrm{BM}=\sqrt{2+\gamma}\,\omega_\textrm{0}$ and $\omega_\textrm{BM}=\sqrt{2}\,\omega_\textrm{0}$, respectively. For interaction potentials proportional to $r^2$, Eq.~\eqref{b}, the confinement has to be of the form $r^2\log(r)$, where the prefactor determines the breathing frequency $\omega_\textrm{BM}$. In case of particle(s) located in the trap origin the confinement is modified containing an additional term $c\,v(r)$, Eq. (8a), or $cr^2$, Eq. (8b). These results are valid for any real $\gamma$ and any dimension and include the result of Ref.~\cite{partoens_etal.jphy97} as a special case.

Further we conclude that no UBM exists for all exponential potentials (such as Yukawa, Morse etc.) or non-monotonic potentials (e.g., Lennard-Jones).

{\bf Non-universal breathing mode (NUBM).} 
A second class of solutions of Eqs.~\eqref{existing_condition}, the configuration dependent solutions, exists in highly symmetric configurations. For instance, 2D equally spaced single shell systems or 3D single shell systems, which form platonic bodies (with or without particle in the center), fulfill Eqs.~\eqref{existing_condition} for any $v$ and $\phi$ -- in these cases the system looks the same from every particle's view. Thus, for every $i\in \{1\ldots N\}$ Eqs.~\eqref{existing_condition} are identical. Then, since no direction is preferred, the sum is proportional to ${\bf r}_i^*$, and a unique value $\lambda\geq0$ exists fulfilling the existence equations. The resulting BM is {\em non-universal}, i.e., $\omega_\textrm{BM}$ depends on $N$.

{\bf Deviations from a BM.}
We will now verify our predictions by a numerical normal mode analysis. To this end, we use the ground state configurations of Lennard-Jones and harmonically confined Yukawa systems for different $N$ \cite{wales97,yukawagroundstate} and calculate all eigenmodes. For each mode we analyze the deviations from a BM, i.e., deviations from the proportionality $\Hat{\bf r}=c\,{\bf r}^*$. The first type of deviations arises if $\Hat{\bf r}_i=c_i{\bf r}^*_i$, with different $c_i$ for different particles (radial, but non-uniform mode). More pronounced deviations exist if there is no proportionality for all particles, i.e., there exist finite tangential velocity components (non-radial mode). Finally, the strongest deviations correspond to the case where the mode is not radial and contains, in addition, a positive number $N_\textrm{ap}$ of anti-phase oscillating particles, i.e., these particles move inward while the majority moves outward and vice versa.

\begin{figure}
\includegraphics[width=7.5714cm,clip=true]{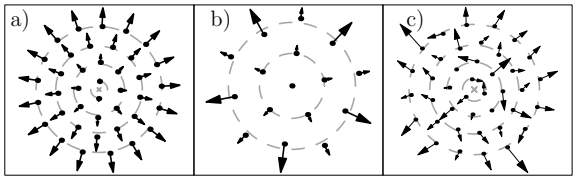}
\caption{QBM of harmonically confined 2D Yukawa systems. a) perfect BM for $N=40$ and $\kappa d_\textrm{c}=0$. b) purely radial but non-uniform QBM for $N=16$ and $\kappa d_\textrm{c}=20.0$, $(\sigma_\textrm{r}=0, \sigma_\textrm{u}=1.1, N_\textrm{ap}=0)$. c) non-radial and non-uniform QBM for $N=40$ and $\kappa d_\textrm{c}=1.99$, $(\sigma_\textrm{r}=2.77, \sigma_\textrm{u}=2.11, N_\textrm{ap}=4)$. The dashed lines mark the shells.}
\label{fig:YukawaResults}
\end{figure}

We will measure these deviations by computing the number $N_\textrm{ap}$ whereas radiality and uniformity will be measured by the distribution widths normalized to the mean
\begin{subequations}
\begin{align}
	\sigma_\textrm{r}&=\left\lvert\max_{i}\delta_{i,\textrm{r}}-\min_{i}\delta_{i,\textrm{r}}\right\rvert/\abs{\bar{\delta}_{i,\textrm{r}}}\textrm{ with }\delta_{i,\textrm{r}}=\frac{\Hat{\bf r}_i\cdot{\bf r}^*_i}{\abs{\Hat{\bf r}_i}\abs{{\bf r}^*_i}}\\
	\sigma_\textrm{u}&=\left(\max_{i}\delta_{i,\textrm{u}}-\min_{i}\delta_{i,\textrm{u}}\right)/\bar{\delta}_{i,\textrm{u}}\textrm{ with }\delta_{i,\textrm{u}}=\frac{\abs{\Hat{\bf r}_i}}{\abs{{\bf r}^*_i}}.
\end{align}
\end{subequations}
If, for each mode, $\sigma_\textrm{r}$ and/or $\sigma_\textrm{u}$ are non-zero, no BM exists. In this case we define the {\em quasi-BM (QBM)} as the mode with the smallest of these deviations. Two examples of such a QBM are shown in Fig.~\ref{fig:YukawaResults} together with a perfect BM.

\begin{figure}
\includegraphics[width=8.5714cm,clip=true]{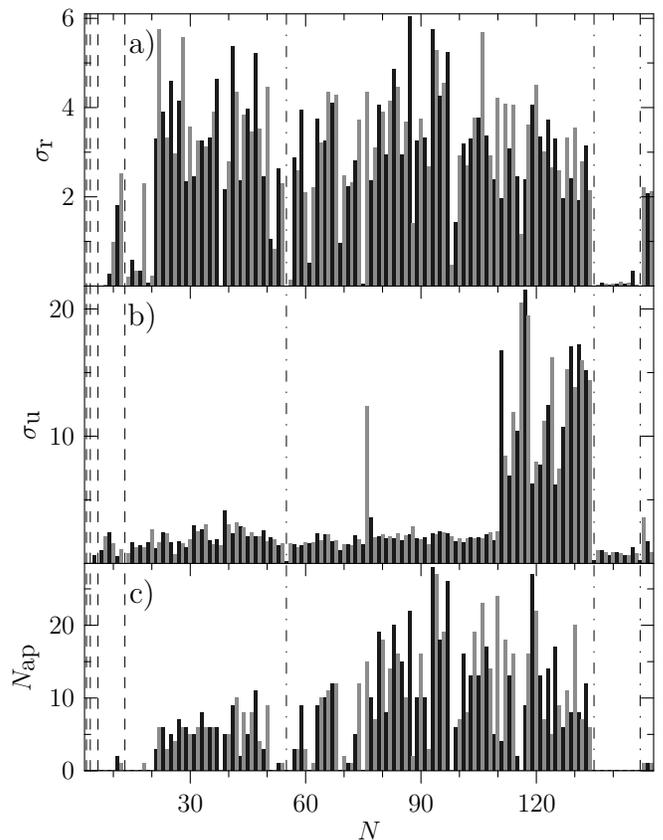}
\caption{QBM of unconfined LJ vs. particle number. a) Deviations $\sigma_\textrm{r}$ from radiality. b) Deviations $\sigma_\textrm{u}$ from uniformity. c) Number $N_\textrm{ap}$ of anti-phase oscillating particles. Cases of NUBM are marked by dashed lines (see also Fig. \ref{fig:nonuniversalBM}) and the QBM of the icosahedral symmetric systems by dashed-dotted lines.}
\label{fig:LJResults}
\end{figure}

{\bf Unconfined 3D Lennard-Jones (LJ) systems.}
First we examine the breathing-type modes of unconfined systems with LJ interaction
\begin{equation}
	v(r)=4\epsilon\biggl[\Bigl(\frac{\sigma}{r}\Bigr)^{12}-\Bigl(\frac{\sigma}{r}\Bigr)^6\biggr],
\end{equation}
which is of relevance, e.g., for molecules and atomic clusters \cite{baletto05,proykova06}. While the eigenvalues of the corresponding modes are dependent on the chosen energy and length scale, the eigenvectors are not. Thus, a systematic calculation of $\sigma_\textrm{r}$ and $\sigma_\textrm{u}$ in dependence on $N$ can be performed. The results are shown in Fig.~\ref{fig:LJResults} where,   $\sigma_\textrm{r}$, $\sigma_\textrm{u}$ and $N_\textrm{ap}$ are plotted vs. $N$ in the range from $3$ to $150$ particles. One clearly sees that most of the systems show substantial deviations from radiality and uniformity. The smallest deviations are observed for $N=55,135,147$, which are highly symmetric multiple shell systems with full icosahedral symmetry \cite{wales97} (see dashed-dotted lines): the values are $\sigma_\textrm{r}=0, N_\textrm{ap}=0$ and $\sigma_\textrm{u}=0.2,\,0.24,\,0.23$, respectively. Thus we confirm that LJ systems do not possess a UBM, and a NUBM exists only for $N=3,4,6,13$ (see dashed lines in Fig.~\ref{fig:LJResults}) and the trivial case $N=2$. These particle numbers correspond to single-shell configurations, which are either planar ($N=3$) or form platonic bodies: tetrahedron ($N=4$), octahedron ($N=6$) and icosahedron with a central particle ($N=13$). Other platonic configurations do not show up.

{\bf Parabolically confined Yukawa systems.}
As a second example of broad practical interest, e.g., for spherical dusty plasmas and colloidal systems, we consider the breathing-type modes in a parabolically confined one-component Yukawa system,
\begin{align}
	v(r)=\frac{Q^2}{r}e^{-\kappa r}, \qquad \phi(r)=\frac{m}{2}\omega_\textrm{0}^2r^2,
\end{align}
using as length scale $d_\textrm{c}=(2Q^2/m\omega_\textrm{0}^2)^{1/3}$ -- the stable distance between two charged particles in the absence of screening \cite{bonitz-etal.prl06}. For $\kappa\rightarrow0$ (Coulomb system), a universal breathing mode is observed, in agreement with the previous analysis, see Fig.~\ref{fig:YukawaResults}.a. In contrast, for $\kappa>0$ our simulations confirm that no UBM exists \cite{sheridan.jphysD06}. Two typical examples of the QBM of 2D Yukawa systems are presented in Fig.~\ref{fig:YukawaResults} clearly showing deviations from radial and uniform motion. While in Fig.~\ref{fig:YukawaResults}.b still a purely radial motion is observed, in Fig.~\ref{fig:YukawaResults}.c, the mode closest to the BM shows striking deviations, including several particles moving in tangential direction.

\begin{figure}
\includegraphics[width=7.5714cm,clip=true]{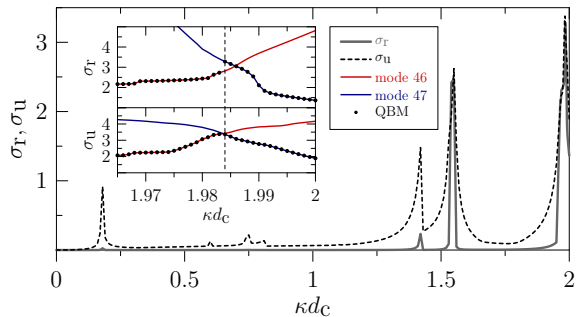}
\caption{(Color online) Deviations $\sigma_\textrm{r}$ (solid line) and $\sigma_\textrm{u}$ (dashed line) of the QBM for harmonically confined 2D Yukawa systems with $N=40$ vs. screening strength $\kappa d_\textrm{c}$. The inset shows $\sigma_\textrm{r}$ and $\sigma_\textrm{u}$ of modes $46$ (red line,  numbers correspond to the mode frequencies in descending order) and $47$ (blue line), which have the smallest deviations. The QBM is presented by dots.}
\label{fig:YukawaResult_N40_k0-k2}
\end{figure}

Let us now analyze the behavior of the QBM as a function of the screening length. Surprisingly, small variation of $\kappa$ leads to an irregular behavior of $\sigma_\textrm{r}$ and/or $\sigma_\textrm{u}$ with very sharp peaks, cf. Fig.~\ref{fig:YukawaResult_N40_k0-k2} showing the properties of the QBM for the example $N=40$ in 2D and $0\le\kappa d_\textrm{c}\le2$.
The origin of the peaks is easily understood: they arise whenever the QBM switches from one eigenmode to another. I.e., one mode having the smallest deviations from the BM in a certain $\kappa$ range loses this role to another mode at some critical value of the screening parameter. This can be seen in the inset of Fig.~\ref{fig:YukawaResult_N40_k0-k2} showing the QBM peak in the region $1.965\le\kappa d_\textrm{c}\le2$.

\begin{figure}
\includegraphics[width=7.cm,clip=true]{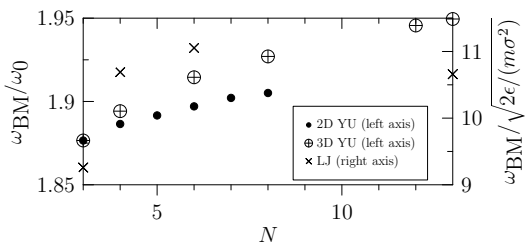}
\vspace*{-0.3cm}
\caption{Frequency of all existing NUBM of 3D LJ (right axis) and 2D and 3D Yukawa (YU) systems with $\kappa d_\textrm{c}=1$ (left axis).}
\label{fig:nonuniversalBM}
\end{figure}

Furthermore, we consider the NUBM of LJ and Yukawa systems more in detail. All existing BM (for a single value of $\kappa$) due to highly symmetric configurations are shown in Fig.~\ref{fig:nonuniversalBM}. Note that, in the case of harmonically trapped Yukawa systems, the values of $N$ exhibiting a NUBM depend on $\kappa$ and on the dimensionality. Further, the simulations confirm that the frequencies are non-universal, i.e., depend on $N$.

Finally, let us now return to the monopole oscillation discussed in the introduction. Considering small harmonic particle oscillations with eigenfrequency $\omega$ around the stationary positions the mean square radius becomes ${\bf R}^2(t)={\bf R}_0^2+2 \sin{\omega t}\, {\bf r}^*\cdot\Hat{\bf r} + O(\Hat{\bf r}^2)$. If the BM exists, the $d\!\cdot\! N-$dimensional vectors $\Hat{\bf r}_\textrm{BM}$ and ${\bf r}^*$ are parallel and there exists a MO with $\omega_\textrm{MO}=\omega_\textrm{BM}$, which is unique, owing to the orthogonality of all eigenvectors. In contrast, if no BM exists, generally many eigenvectors $\Hat{\bf r}^{(k)}$ with frequency $\omega_k$ will have a non-vanishing projection on ${\bf r}^*$, each giving rise to a MO with different frequencies $\omega_\textrm{MO}^{(k)}=\omega_k$ and amplitudes. In this case, an external perturbation of the system which excites radial particle motions will excite all those monopole oscillations at once. The resulting complex spectrum is a fingerprint of the internal properties of the system, in particular the form of the pair interaction and the confinement, and is expected to be a sensitive experimental diagnostic of strongly correlated systems.

In summary, we have derived the general existence conditions for the BM in strongly correlated finite classical systems of any dimension with arbitrary confinement and interaction potential. The existence of a BM is restricted to a small class of systems: i) parabolically confined systems with a power law or logarithmic interaction potential, or ii) particular, highly symmetric configurations, independently of the interaction. Systems with exponential or non-monotonic potentials (including Yukawa, Morse or Lennard-Jones systems) not having such a special configuration do not possess a BM. In these systems all normal modes exhibit non-uniform and/or non-radial motion which has been illustrated for several representative examples. 

This work is supported by the Deutsche Forschungsgemeinschaft via SFB-TR 24 and the German Academic Exchange Service via the RISE program.

%

\end{document}